\documentclass[12pt,aps,preprint]{JHEP3}
\usepackage[latin1]{inputenc}
\usepackage{bm}
\usepackage{amsmath,amssymb,graphicx,mathrsfs,epsfig}

\renewcommand{\a}{\alpha}

\newcommand{\eps}{\epsilon}
\newcommand{\p}{\phi}

\newcommand{\be}{\begin{equation}}
\newcommand{\ee}{\end{equation}}
\newcommand{\bea}{\begin{eqnarray}}
\newcommand{\eea}{\end{eqnarray}}

\title{ Cosmic Microwave Background observables of small field models of inflation}
\author{Ido Ben-Dayan, Ram Brustein\\
Department of Physics, Ben-Gurion University,
    Beer-Sheva 84105, Israel \\ E-mail: idobd@bgu.ac.il, ramyb@bgu.ac.il}
\preprint{}
\abstract{
We construct a class of single small field models of inflation that can predict, contrary to popular wisdom, an observable gravitational wave signal in the cosmic microwave background anisotropies. The spectral index, its running, the tensor to scalar ratio and the number of e-folds can cover all the parameter space currently allowed by cosmological observations. A unique feature of models in this class is  their ability to predict a negative spectral index running in accordance with recent cosmic microwave background observations. We discuss the new class of models from an effective field theory perspective and show that if the dimensionless trilinear coupling is small, as required for consistency, then the observed spectral index running implies a high scale of inflation and hence an observable gravitational wave signal. All the models share a distinct prediction of higher power  at smaller scales, making them easy targets for detection.
}

\begin{document}

\section{Introduction}

Inflation, a period of accelerated expansion of the early universe, is an accepted part of the standard cosmological model. Viewed from a fundamental physics perspective inflation can occur in a complicated multi-dimensional field space. However, in most simple
cases it is possible to identify, at least a-posteriori, a single inflaton and hence it is possible to use effective single field dynamics. Single field models of inflation are conveniently classified into two main classes, large field models and small field models. The potential of the latter contain only a small region in field space where the slow roll conditions are valid. As effective field theories small field models are easier to realize, since the slow roll conditions can be satisfied accidentally in a small region near a maximum or a saddle point with a small potential curvature and the higher order terms are generically suppressed, as opposed to large field models that require controlling the effective potential over large regions in field space. Small field models require the least input from the microscopic physics that generates inflation and are thus the most economic parametrization when this physics is not known in detail.

Popular wisdom says that, as we explain below, small field models of inflation cannot predict an observable signal of gravitational waves (GW) in the cosmic microwave background (CMB) \cite{lyththrm,Lyth:1998xn,McAllister:2007bg,Kallosh:2007ig,Challinor:2009tp}. Additionally, simple small field models of inflation predict a red spectrum of scalar perturbations and negligible spectral index running (RUN) \cite{BenDayan:2008dv} and for most analytical models significant RUN requires a significant departure from slow-roll conditions \cite{Joy:2008qd,Hunt:2004vt,Hunt:2007dn}.

The QUaD experiment has recently reported a 2$\sigma$ detection of negative RUN \cite{quad} which if verified will imply that the simplest small field models are disfavored. The result is based on combining the CMB observations of the QUaD, WMAP5 \cite{WMAP5} and ACBAR \cite{acbar} experiments with the large scale structure observations of the SDSS \cite{SDSS}. The very recent WMAP7 result \cite{WMAP7} by itself is just under the 2$\sigma$ mark.
So far only upper bounds have been set on  GW.  The upper bounds on the GW component are given in terms of the tensor to scalar fluctuations ratio $r$.  Depending on prior assumptions about the spectral index $n_S$ and its running $\alpha=\frac{d n_S}{d \ln k}$ the bounds are as large as $r<0.55$ or as small as $r<0.2$ \cite{quad,WMAP5,WMAP7}. Recently,  the first direct upper bound from B-mode polarization was published \cite{Chiang:2009xs} and the bounds are expected to be improved significantly in the future \cite{planck}, perhaps reaching a value as low as $r\sim 0.01$. As we have mentioned the current value for $\alpha$ is $-.052\pm.023$ reported by the QUaD experiment \cite{quad}. Future observations are likely to significantly improve the errors on the value of $\alpha$. For example, after one year of data the PLANCK mission alone will measure the running at the level of $\Delta \a \approx .005$ (if the running is scale independent).

It is rather disappointing that the small field models that are such an interesting class of models from a theoretical perspective  is seemingly disfavored by observations and furthermore predicts that observers will come up empty handed when trying to detect GW, which is a major goal of the current and future CMB observations. We will show that, contrary to popular wisdom, interesting small field models can predict observable GW  and observable negative running and that these predictions are related to the fact that the rate of change of the Hubble parameter during the era when most e-folds were accumulated can be smaller than its value at the CMB point.

The paper is organized as follows. In the next section we present the new class of models and their predictions for the CMB, we then discuss the models from an effective field theory perspective. We end with a brief summary and conclusions.

\section{A new class of inflation models}

A simple argument shows that to produce observable GW in single field models of inflation, the inflaton $\phi$ has to  move a large distance in field space in reduced Planck scale units.  Our conventions are such that $m_p\equiv\frac{1}{\sqrt{8\pi G_N}}=2.4\times 10^{18} GeV$. The scalar to tensor ratio is given in terms of the slow-roll (SR) parameter $\epsilon=1/2 \left({\dot \phi}/H\right)^2\simeq 1/2 \left(V'/V\right)^2$ as $r=16\epsilon$. The number of e-folds $N$ can be expressed in terms of $r$:
$
\frac{dN}{d \phi}=\sqrt{\frac{8}{r}}
$.
If $\epsilon$ is approximately constant during the last $N_{CMB}\sim 60$ e-folds of inflation then
$
r\simeq 8 \left(\frac{\Delta \phi}{N_{CMB}}\right)^2
$.
The argument was presented in a detailed form by Lyth \cite{lyththrm,Lyth:1998xn}. While the ``Lyth-theorem" states only that a field motion of about $m_p$ is sufficient, in practice, inflationary models that produce observable GW actually require that the inflaton moves about $10m_p$ or more \cite{Linde:1983gd,Silverstein:2008sg,McAllister:2008hb,Cicoli:2008gp,Hotchkiss:2008sa} leading to the wide-spread belief that discovering GW will rule out all small field models \footnote{See, for example, Eq.(94) in \cite{Challinor:2009tp}}.

From the previous argument it would seem that  $r \sim 0.01$ with about 60 e-folds of inflation requires $\Delta \phi \sim 2$,  so why hasn't such a simple ``small field model" been constructed? Recall that $\epsilon$ cannot be a true constant during inflation, since inflation ends only when $\epsilon=1$. In realistic models, to conform with observations, the SR parameter  $\eta\sim d\epsilon/d\phi$  has to be quite small as well and this constraint forces the range of motion to be larger.  Similar consideration show that in realistic models the observational bounds on the  running of the spectral index also significantly increase the range of motion. Additionally, in most known constructions (See, for example, \cite{Lyth:1998xn}) $\epsilon$ is always monotonic.

While the SR parameter $\epsilon$ is bounded $0<\epsilon<1$, it does not need to be constant or monotonic in any fundamental way. Rather one typically has to tune less parameters if $\epsilon$ is constant or monotonic. For designing models that produce an observable GW signal it is enough to add a single parameter to the models in \cite{BenDayan:2008dv}, $V(\p)=1-a_1\phi-a_2\p^2-a_p\p^p$. We have chosen a more general Taylor expansion that
allows for a varying $\epsilon$. We will show later that such a choice is quite natural from an effective field theory perspective. The idea is to design models that interpolate between values corresponding to large field models of about $\sqrt{2\epsilon} \sim 0.1$ and values corresponding to small field models $\sqrt{2 \epsilon}<0.01$. The sufficient number of e-folds is obtained from the region where $\epsilon$ is small and the large enough tensor amplitude is obtained from the region where $\epsilon$ is relatively large. The relatively large changes in $\sqrt{2\epsilon}$ over a small region in field space could mean that $V'''/V$ has to be relatively large and consequently that the RUN $\alpha$ may be large enough to be observable.
Our idea is illustrated in Fig.~1 where three models are shown. One of the models is a large field monomial model (green), another is a  small field SUGRA model of the class presented in \cite{BenDayan:2008dv} (blue), and the third is an interpolating  model from our new class (red).
\FIGURE[t]{
\scalebox{0.9}{\includegraphics{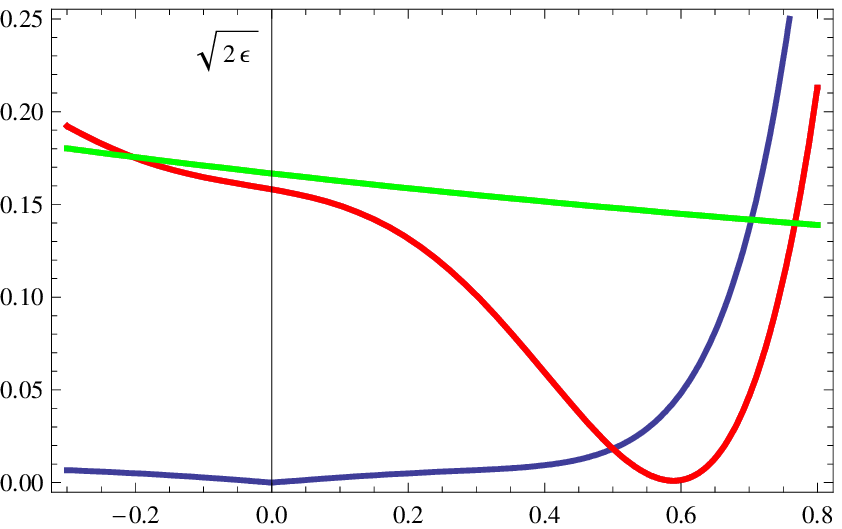}}\vspace{.1in}  \scalebox{1.01}{\includegraphics{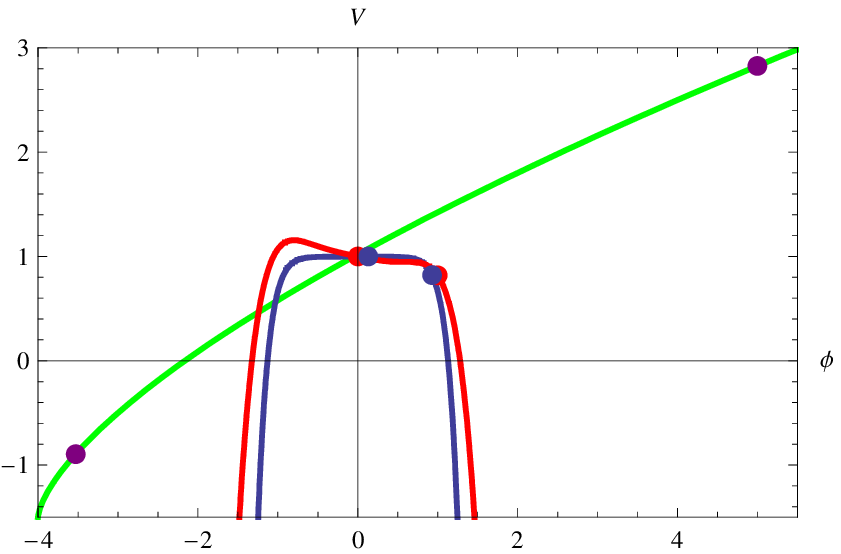}}
\caption{ \label{figure1} Shown is a graph of $\sqrt{2 \epsilon}=V'/V$ (upper panel) and $V$ (lower panel)  for a small field canonical SUGRA model (blue), a large field model (green) and a model of the new class with non-monotonic $\epsilon$ (red). The new model interpolates between the two others. For the small field model (blue) the CMB point is at $\phi_{CMB}=0.13$ and inflation ends at $\phi_{END}=0.93$. For the large field model (green) $(\phi_{CMB}=5,\phi_{END}=-3.53)$ and for the new model (red) $(\phi_{CMB}=0,\phi_{END}=1.0)$. The large field model is offset $V\rightarrow V-1.5$. Additionally, to demonstrate the similarity between the small field model(blue) and the new model (red)  a symmetric example was chosen, i.e. $a_5=0, a_6=0.3911$. The CMB observables are $n_s=1.01, r=0.23, \a=-0.04$.
}}
%\end{figure}

In the new class of models we use five parameters; three slow roll parameters, the number of e-folds $N_{CMB}$ from the CMB era to the end of inflation and the value of the inflaton $\phi_{END}$ at the end of inflation.
The potential $V(\p)$ is expressed in terms of the three SR parameters
$
\epsilon=\frac{1}{2}\left(\frac{V'}{V}\right)^2
$,
$
\eta=\frac{V''}{V}
$, and
$
\xi^2=\frac{V^{'''}V'}{V^2}
$ as follows:
\begin{equation}
\label{poten}
\frac{V(\p)}{V(0)}=1 - \sqrt{\frac{r_0}{8}}\p + \frac{\eta_0}{2} \p^2 + \frac{\a_0}{
 3 \sqrt{2r_0}}\p^3 - a_4 \p^4 - a_5 \p^5.
\end{equation}
As their names suggest, $r_0,\eta_0$ are the desired values of these observable/parameter at the CMB point, $(\p\!=\!0)$, and $\a_0=-2\xi^2(\p=0)$.
The value of the field at the end of inflation when $\epsilon(\phi_{END})=1$ determines $a_4$,
\be
 \frac{1}{2}\left(\frac{-\sqrt{\frac{r_0}{8}} + \eta_0 \p_{END} + \frac{\a_0}{
 \sqrt{2r_0}}\p_{END}^2 - 4 a_4 \p_{END}^3 - 5 a_5 \p_{END}^4}{1 - \sqrt{\frac{r_0}{8}}\p_{END} + \frac{\eta_0}{2} \p_{END}^2 + \frac{\a_0}{
 3 \sqrt{2r_0}}\p_{END}^3 - a_4 \p_{END}^4 - a_5 \p_{END}^5}\right)^2=1
\ee
and the number of e-folds determines $a_5$,
\begin{equation}
N_{CMB} = \int_0^{\p_{END}}\frac{d\p}{\sqrt{2 \eps(\p;a_5)}}
.
\end{equation}
Obviously, any other polynomial potential with five parameters will do as well.

That any model that satisfies all the requirements exists is not at all obvious from the previous discussion because of the two apparently highly non-linear constraints that need to be satisfied: that the field moves only a Planck distance in field space and that the total number of e-folds is larger than about 60. However, if the potential is expanded around the point $\phi_{min}$ where $\sqrt{\epsilon}$ has a quadratic minimum it becomes clear that many solutions exist. There the potential has the expansion $V(\p)/V(\phi_{min})= 1+b_1 (\phi-\phi_{min}) + b_3 (\phi-\phi_{min})^3+\cdots$ where the dots stand for higher order terms.  Then it is clear that $b_1$ and $b_3$ determine in a simple way the number e-folds that are accumulated near $\phi_{min}$ and the range of motion of $\phi$. The higher order terms that are essentially irrelevant near $\phi_{min}$ can then be used to tune the CMB observables to any desired value without affecting much the dynamics near $\phi_{min}$.

In our models the third derivative of the potential can be large enough to influence the value of the spectral index, hence the corresponding CMB observables are evaluated to the next order in the SR approximation are \cite{Lyth:1998xn}\footnote{To agree with $\alpha=dn_s/d\ln k$ our sign differs from the one in \cite{Lyth:1998xn}.}:
\bea
n_s&=&1+2\eta-6\eps+2\left[\frac{1}{3}\eta^2+(8C-1)\eps \eta-\left(\frac{5}{3}-C\right)\eps^2-\left(C-\frac{1}{3}\right)\xi^2\right]\\
%$n_s=1+2\eta-6\eps+\left[3(5c-3)\eps \eta-\frac{1}{2}\left(1+9c\right)\eps^2+\frac{1}{6}\left(13-3c\right)\xi^2+\frac{1}{6}(5c-3)\eps\xi^2\right]$,
r &=&16 \epsilon\\
\alpha &=&16 \epsilon \eta-24\epsilon^2-2\xi^2,
 \eea
where $C=2+2\ln 2+\gamma=-0.73$, $\gamma$ being the Euler constant.  These expressions are based on approximating the slow-roll parameters as constants in the perturbation equation. In some models, this approximation proved not to be accurate enough to correctly determine the observables.
Hence, to compute the spectrum we have solved numerically the full Mukhanov-Sasaki equation
\be
\frac{d^2 u_k}{d\tau^2}+\left(k^2-\frac{1}{z}\frac{d^2z}{d\tau^2}\right)u_k=0,
\ee
where the $d\tau=dt/a(t)$ is the conformal time and $z=a\dot{\p}/H$.
The power spectrum is then given by
\be
P_R(k)=\frac{k^3}{2\pi^2}\left|\frac{u_k}{z}\right|^2.
\ee
The observables were extracted from the power spectrum by choosing the suitable pivot scale.
In the case of significant running the higher order terms induce some level of running of running in accord with \cite{eastherpeiris}.
The PLANCK satellite is sensitive to $\Delta \alpha=0.005$, assuming constant running. If PLANCK  is sensitive enough to detect running of running it will probably be better to use the entire power spectrum for comparison with the data rather than comparing the parameters $n_s$ and $\alpha$.

The non-monotonic $\epsilon$ which is used to generate significant GW signal also leads to the most unique property of the models which makes them distinguishable from other models with similar observables. The field spends only a small number of e-folds near the CMB scale where the SR parameter $\eps(0)$ is larger and rolls down to a point where $\eps$ is minimal where most e-folds are generated. Considering that the power spectrum $P\sim \frac{V}{\eps}$, we see that there should be a significant enhancement of the spectrum at smaller scales. This enhancement is not limited to the CMB scales and hence could be, in principle, tested in the matter power spectrum. An accentuated example of both phenomena (running of running and enhancement at smaller scales) is presented in Figure~\ref{figure2}.

\FIGURE[t]{
\scalebox{0.9}{\includegraphics{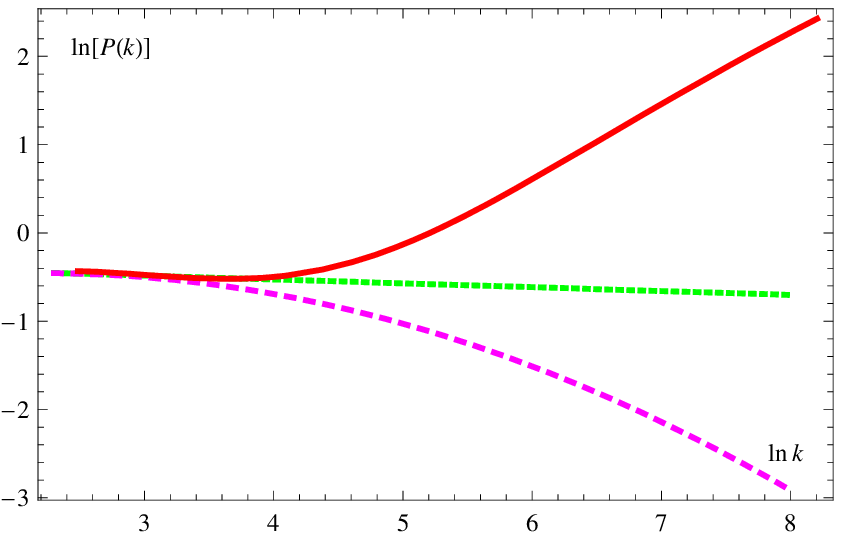}}\vspace{.1in} \scalebox{.94} {\includegraphics{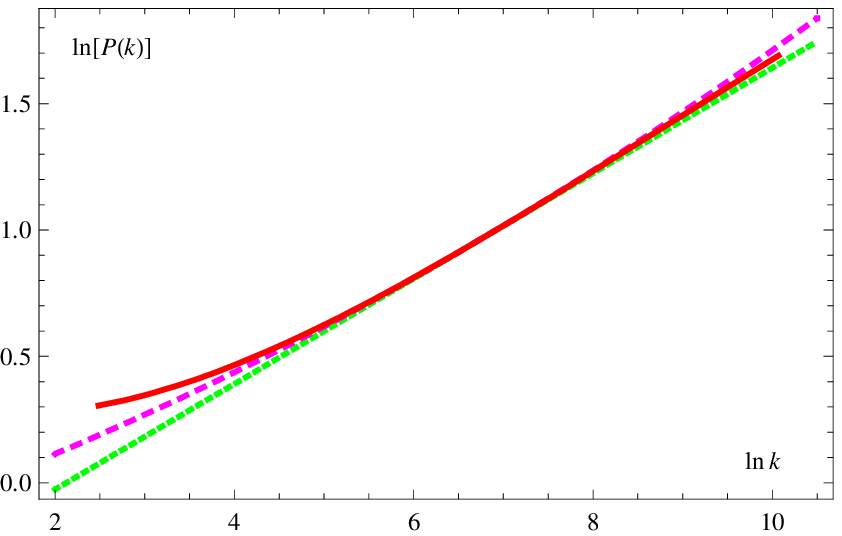}}
\caption{ \label{figure2} Shown is a plot of  $\ln P(k)$ vs. $\ln k$ for the first model (upper panel) and the last model (lower panel) in Table~\ref{table:models}. The observables in the table are evaluated at a pivot scale such that $\ln (k/k_0)=2.5$ (upper panel) and $\ln (k/k_0)=6.7$ (lower panel). The green dotted curves assume that the spectrum is approximated by a constant spectral index. The pink dashed curves assume that the spectrum is approximated by the values of the spectral index and the running at $k_0$:  $n_S(k_0), \alpha(k_0)$.}
}
%\end{figure}
In Figure~\ref{figure2} we plot the logarithm of the power spectrum vs. $\ln k$ for the observable scales of a model with significant running. This depicts the sensitivity of the traditional observables $n_S$ and $\alpha$ to the pivot scale $k_0$. This also shows the way to test these models since over a large enough range of scales the model predictions deviates from those of a model with a constant $n_S$ and a constant $\alpha$.

\begin{table}[t]
%\TABLE[t]{
\begin{center}
\hbox{\hspace{1.2in}Potential parameters\hspace{1.1in}Range\hspace{.3in} CMB observables}\vspace{.05in}
\begin{tabular}{|c|r|r|r|l||c|c||c|c|r|}
  \hline
   \ $r_0$ & $\eta_0$ & $\a_0$ & $a_4$ &  $a_5$ & $\ \Delta\phi_{50}\ $ & $\ \Delta\phi_{60}\ $ & $n_s$ & $r$ & $\a$  \\
    \hline
\ $* \ \S \ 0.10$ & $ 0.015 $ & $0.03$ & $-0.6102$ & $ 0.709 $ & $0.57$ & $1.0$ & $0.96$ & $0.10$ & $-0.07$  \\
   \hline
   \ $\S \  0.02$ & $ 0.01 $ & $ 0.005 $ & $-0.875$ & $ 1.451 $ & $0.40$ & $0.8$ & $0.99$ & $0.02$ & $0.001$  \\
   \hline
   \ $\S \ 0.08$ & $ -0.005 $ & $ -0.02 $ & $-0.695$ & $ 0.7567 $ & $ 0.57$ & $1.0$ & $0.97$ & $0.08$ & $-0.05$ \\
   \hline
   \ $\S \ 0.10$ & $ 0 $ & $ 0 $ & $-0.688$ & $ 0.7591 $ & $ 0.57$ & $1.0$ & $1.01$ & $0.10$ & $-0.006$ \\
   \hline
\ $\S \ 0.05$ & $ -0.02 $ & $ -0.03 $ & $-0.6834$ & $ 0.7405 $ & $ 0.56$ & $1.0$ & $0.94$ & $0.05$ & $-0.06$   \\
\hline
\ $\ 0.01$ & $ 0 $ & $ 0 $ & $-0.3919$ & $ 0.538 $ & $ 0.49$ & $1.0$ & $0.99$ & $0.01$ & $0.001$   \\
\hline
\ $*\ 0.02 $ & $ 0.108 $ & $ 0.003 $ & $0.0341$ & $\ \  0 $ & $ 0.80$ & $2.0$ & $1.21$ & $0.02$ & $0.005$ \\
   \hline
   \end{tabular}
\end{center}
\caption{Listed are the values of the potential parameters, the range of inflaton motion after 50 and 60 e-folds
and the values of the CMB observables, assuming that $N_{CMB}=60$.
The models appearing in Figure 2 are marked with an asterisk. The models appearing in Figure 3 are marked with $\S$. The last model is a renormalizable model with $a_5=0$.}
\label{table:models}
% .}
\end{table}
To demonstrate the richness of this class of models we present several specific examples with a wide range of CMB observables in Table~\ref{table:models}. Some of the models in Table~\ref{table:models} are already ruled out by the QUaD results and others can be confirmed (or ruled out) by upcoming experiments, in particular, the PLANCK satellite.  In all the examples the CMB point is at $\p=0$ and  $N_{CMB}=60$.
%The main region of parameter space where one can find only models of the new class and none of the traditional models is in the region of large negative running.

\FIGURE[t]{
\hspace{-1.1in}\scalebox{0.5}{\includegraphics{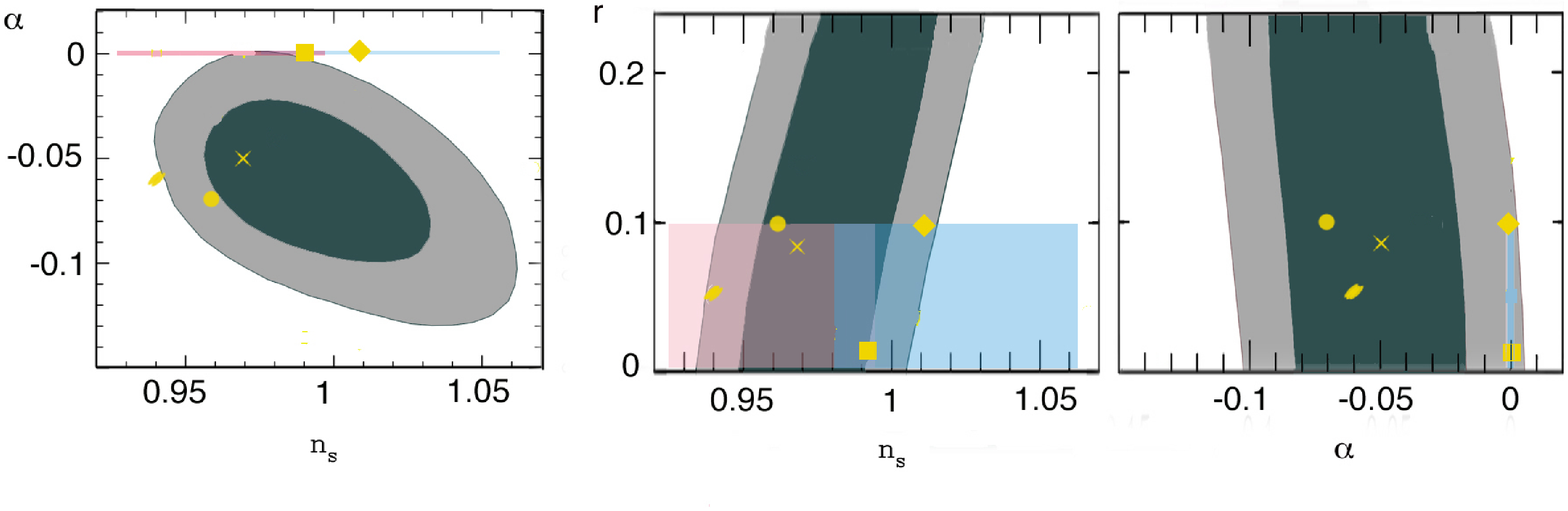}}\hspace{-1.3in}\vspace{-.7in}
\caption{ \label{figure3} Shown are model predictions for models of Table~\ref{table:models} marked with a $\S$ for various CMB observables on the background of the QUaD analysis of their CMB allowed region. The center and right panels show $r$ vs. $n_s$ and $\alpha=d n_s/d\ln k$ (respectively). The left panel shows $\alpha$ vs. $n_s$. The pink and blue rectangles depict the regions of parameter space that traditional models occupy as explained in the text.}
}

In Figure~\ref{figure3} we present the CMB observables of several of the new models overlaid on the QUaD results. The models are compared to traditional large field models that are represented by the pink rectangle and the traditional hybrid models that are represented by the blue rectangle. The choice $r \le .1$ for these models is somewhat arbitrary and represents the range for $r$ that they typically predict.  Note that the pink and blue rectangles overlap in the center panel of Fig.~\ref{figure3}. In the right panel of Fig.~\ref{figure3} both large and hybrid models are represented by the thin vertical line centered at $\alpha=0$.  The  traditional small field models are not shown at all in Fig.~\ref{figure3} because they predict negligibly small $r$ and $\alpha$. The main region of parameter space where one can find only models of the new class and none of the traditional models is in the region of large negative running. This is shown on the left and right panels of Fig.~\ref{figure3} where the predictions of the traditional models are depicted by thin lines (horizontal or vertical, respectively) centered at $\alpha=0$.

\section{Effective field theory considerations}

Let us now discuss our new class of models from an effective field theoretical (EFT) perspective. Since at the moment we do not have a microscopic theory that predicts the identity of the inflaton and its couplings we have to limit ourselves to macroscopic considerations.  As explained in detail in \cite{Burgess:2009ea},
the potential is expected to have an expansion in powers of $\phi$  with coefficients whose values are determined by dimensional analysis:
\be
V=\Lambda^4\left(1+\sum\limits_{n=1}\lambda_n (\phi/m_P)^n \right).
 \ee
Here we have incorporated the information that for our models the potential remains roughly constant when the field varies by about a Planck distance. The scale $\Lambda=V^{1/4}$ is determined by the standard argument $\Lambda\simeq 1\times 10^{16}\ \text{GeV}\ (r/.01)^{1/4}$ (see, for example, \cite{Challinor:2009tp,German:1999gi}), so for  $r>.01$ a very high scale for inflationary models is expected. A necessary condition for the validity of any EFT is that the marginal and irrelevant dimensionless couplings are small $\lambda_n\ll 1$, $n\ge 4$, while the superrenormalizable terms with $n\le 3$ have to be treated separately.  From observations we know that the first three coefficients in the potential are constrained  to be quite small, for example, the first coefficient is $\lambda_1=-.035 (r/.01)^{1/2}$. However, the higher order coefficients can be larger, as in the models of Table~\ref{table:models}.
The trilinear coupling $\lambda_3$ is of particular interest to us. The important relevant fact about it is that it can induce interaction vertices that depend on negative powers of the probed energy scale  $\sim(\Lambda/E)$. As the detailed discussion of \cite{Burgess:2009ea} concludes, this means that for the EFT description to be valid for energy scales above the Hubble scale $\Lambda\gtrsim E\gtrsim H$ the trilinear coupling has to be small $\lambda_3\ll 1$. We will also argue shortly, that the fundamental scale of the theory and $\Lambda$ are not too far from each other. These observations are central to the following.

The fundamental scale of the microscopic theory $\Lambda_{UV}$ that produces the inflaton potential is  expected to be somewhat below the (reduced) Planck scale. For example, in weakly coupled string theory the scale is the string mass scale $M_s=M_P g_s$ which is related to the (unreduced) Planck scale via the small string coupling $g_s$. With $g_s$ of the order of the unified gauge coupling this leads to $M_s \sim\text{a few}\times 10^{17} GeV$. For moderately or strongly coupled string theory estimates of the potential scale \cite{Brustein:2002mp} are in the range of $\text{a few}\times 10^{16} GeV$. The main point that we wish to emphasize is that we expect the fundamental scale to be {\it at most} as high as $\text{a few}\times 10^{17} GeV$ and perhaps even as low as $\text{a few}\times 10^{16} GeV$. Obviously, if the fundamental scale is lower than $~10^{16} GeV$, the microscopic theory cannot induce an inflaton potential with scales that are high enough to produce observable GW.
The conclusion from the previous discussion is that for models that produce observable GW signal in the CMB the potential energy scale $\Lambda$ is quite close to the fundamental scale of the microscopic theory $\Lambda_{UV}$. Because the separation of scales $\Lambda/\Lambda_{UV}$ is relatively small the standard arguments about renormalizability, scales and flow of couplings in the infrared can only serve as an approximate guide to estimating the size of potential parameters.

The specific class of models that we have considered generically contain a dimension 5 operator $~\phi^5$ in the potential in addition to relevant and marginal terms.  As we have just argued, for the case of small scale separation considering a dimension 5 operators (and perhaps also dimension 6) is natural. However, because the scales are nevertheless separated  we can expect only a finite small number of potential terms to be important for the inflaton dynamics.   Additionally, the small scale separation means that the parameters of the EFT are sensitive to the microscopic details and can vary from model to model.

We argue that for our class of models a RUN  $|\a|\sim.05$ is an indication of a high scale of inflation and hence of a large GW signal. In other words, for models that produce a RUN of about $.05$ the value of $r$ has to be large for consistency. For models that produce RUN of  the QUaD magnitude we may use the approximate form for $\a$,
\be
|\a|\simeq 2 |\xi^2| =2 m_p^4 \frac{|V''' V'|}{V^2}.
 \ee
 Since $m_p^3\frac{|V'''|}{V}=3!\ |\lambda_3|$ and since $m_p\frac{|V'|}{V}=\sqrt{r/8}$ we obtain
 \be
 r=2(|\a/(3! \lambda_3)|)^2.
\ee
Let us now define $r_{0.01}\equiv r/0.01$ and $\a_{0.05}\equiv |\a|/0.05$ and $|\widehat\lambda_3|\equiv 3! |\lambda_3|$. Then $r_{0.01}=.5\ |\a_{0.05}^2 {\widehat\lambda_3}^{-2}|$. Imposing the condition that we have established previously for the validity of the EFT $|\lambda_3|\ll 1$, leads to a lower bound on the GW strength
\be
r_{0.01}>.5\ \a_{0.05}^2,
\ee
which implies that if the value of the RUN is the one observed by QUaD then we should expect an observable $r$.
Similarly, using the standard estimate $\Lambda\simeq 1\times 10^{16}\ \text{GeV}\ (r_{0.01})^{1/4}$ we obtain
\be
\Lambda \simeq 8.5 \times 10^{15}\ \text{GeV}\ \left(|\frac{\a_{0.05}}{\widehat\lambda_3}|\right)^{1/2}
\ee
which leads to a lower bound on the scale of inflation
\be
\Lambda> 8.5 \times 10^{15}\ \text{GeV}\ (|\a_{0.05}|)^{1/2}.
\ee

\section{Conclusions}

In this paper we have constructed a new class of inflationary models. The unique property of the new class is its ability to predict detectable gravitational waves (GW) and detectable running spectral index (RUN) while the field excursion in Planck units is still small, $\Delta \phi\leq 1$. Moreover, using effective field theory considerations, we have shown that these observables are related and a detection of RUN yields significant lower bounds on the scale of inflation and the GW signal. Our approach is  complementary to the approach  usually used in the literature where one a-priori constrains the number of free parameters in the model and gets connection between observables. Another generic feature of this new class is the higher power at smaller scales. As explained in the text this comes from the non-monotonic nature of the slow-roll parameter $\epsilon$. This unique prediction will enable observations to distinguish between the new class and previous models with large GW or RUN. It is clear from the above analysis that the running is a better discriminator between models than the conventional $r$ and $n_S$. If running is actually detected, then this work as well as \cite{eastherpeiris} imply that the simplest slow-roll picture is insufficient. In this sense we believe that it is better to use the entire spectrum to confirm or rule-out models rather than the standard $n_S$ (or $n_S$ and $\alpha$) power-law parametrization.

A unique and important diagnostic in this context is non-gaussianity in the CMB anisotropies. Naively, the enhanced running in our class of models could have yielded a large non-gaussianity signal. Nevertheless, as stated in \cite{Maldacena:2002vr} (see also \cite{wein}) single field models yield  a non-gaussianity parameter $f_{NL}<1$ while the PLANCK satellite will be able to detect $f_{NL}>5$. Hence such a measurement will be extremely useful for deciding whether inflation can be effectively described by single field models or not, perhaps forcing us to consider a more complicated paradigm for inflation. We intend to address this issue in detail soon.

\section{Acknowledgements}
We thank R.~Bond,  Z.~Huang, M.~Nolta, L.~Page, J.~Sievers ,  D.~Spergel and P.~Steinhard   for useful explanations, comments and suggestions. We thank N.~Itzhaki in particular for many useful discussions. We thank Michael Brown and the QUaD collaboration for providing us with published and unpublished material. This research is supported by ISF grant 470/06. IBD would like to thank the Hebrew University in Jerusalem for their hospitality for the duration of the research.

\end{document}